# Research Article
# Trading Strategies of a Leveraged ETF in a Continuous Double Auction Market Using an Agent-Based Simulation

Isao Yagi ,[1] Shunya Maruyama,[2] and Takanobu Mizuta[3]

[1]*Faculty of Information Technology, Kanagawa Institute of Technology, Atsugi 243-0292, Japan*
[2]*Course of Information and Computer Sciences, Graduate School of Kanagawa Institute of Technology, Atsugi 243-0292, Japan*
[3]*SPARX Asset Management Co., Ltd., Minato-ku 108-0075, Japan*

Correspondence should be addressed to Isao Yagi; iyagi2005@gmail.com

A leveraged ETF is a fund aimed at achieving a rate of return several times greater than that of the underlying asset such as Nikkei 225 futures. Recently, it has been suggested that rebalancing trades of a leveraged ETF may destabilize the financial markets. An empirical study using an agent-based simulation indicated that a rebalancing trade strategy could affect the price formation of an underlying asset market. However, no leveraged ETF trading method for suppressing the increase in volatility as much as possible has yet been proposed. In this paper, we compare different strategies of trading for a proposed trading model and report the results of our investigation regarding how best to suppress an increase in market volatility. As a result, it was found that as the minimum number of orders in a rebalancing trade increases, the impact on the market price formation decreases.

## 1. Introduction

A leveraged exchange-traded fund (ETF) is a fund aimed at achieving a rate of return several times greater than that of the underlying asset, such as Nikkei 225 and S&P futures. These kinds of leveraged ETFs invest in financial market indexes, so they are generally less risky than individual stocks. However, through their use of leveraging, they are able to provide high returns, so the volume of assets under their management has dramatically expanded. Therefore, the risk of a leveraged ETF is also high. Note that we cannot necessarily say that the risk of a leveraged ETF is higher than that of individual stocks. Leveraged ETFs have to execute rebalancing trades on a daily basis—that is, buy the underlying asset when its price goes up and sell it when it goes down—in order to maintain their level of leverage. Therefore, it has been suggested that rebalancing trades of a leveraged ETF may cause destabilization of the prices of the underlying assets [1].

There have been some empirical studies reporting that a leveraged ETF affects the underlying asset markets [2–6]. Chen and Madhavan [2] suggested that by purchasing assets following positive returns and selling assets following negative returns, leveraged ETFs exert additional upward price pressure on the underlying assets following positive returns and additional downward pressure following negative returns, both of which amplify market movements. Rompotis [5] examined how UK leveraged ETFs affected their underlying indexes and found that they could perform their daily rebalancing trades correctly but tracking errors—which refers to the divergence between the daily return of the leveraged ETF and the leverage ratio of the ETF times the corresponding return of underlying asset—became larger as market volatility increased. However, it is difficult for us to determine whether these findings are the pure contribution of the rebalancing trades of the leveraged ETFs to the underlying asset markets, since many factors can affect price formation in actual markets.

One way of analysing how particular transactions influence the financial market is to use an artificial market [7–9]. An artificial market is a multiagent-based model of financial markets. Each of the agents is assigned a specific trading (i.e., buying and selling) rule and then set to trade financial assets as an investor. The market can then be observed to see how the agents behave. At the same time, it is possible to examine what kinds of effects their behaviours induce in the market.

As described in our previous paper [10], artificial markets have been proposed and refined by many researchers. For example, Chiarella et al. [7] succeeded in structuring a simple agent model which reproduced the statistical characteristics of the kinds of long-term price fluctuations observed in empirical analyses. Chen et al. [8] suggested that a simple a model which can reproduce the desired stylized facts should be made as possible.

Studies on artificial markets have had some success in market analysis, such as investigation of the impact of market regulations on the market [11–13]. In particular, Yagi and Mizuta [11] observed the correlation between the amount of leveraged ETF trading and the volatility of the underlying market and found that the rebalancing trades can destroy the underlying asset's market if the rebalancing trade impact on the market is greater than that of ordinary volatility of the underlying assets.

However, there have been no proposals of a leveraged ETF trading method that suppresses the increase in volatility as much as possible. In actual markets, it is up to the fund managers to decide whether to carry out the rebalancing trades at once or disperse them throughout the day. According to Yagi and Mizuta [11], since a larger amount of the managed assets of leveraged ETFs corresponds to a higher volatility of the underlying asset markets, it seems preferable to perform the rebalancing trades dispersed through the day.

In this paper, we consider a rebalancing trade method that has as little influence on the underlying market volatility as possible while maintaining small tracking errors by implementing a leveraged ETF trading model in the artificial market developed by Yagi et al. [14]. The amount of the rebalancing trading orders depends on the size of the tracking error. In our study, the rebalancing trade orders are placed when the amount of the rebalancing orders is larger than a preset threshold, which means the minimum amount of orders that are actually placed as the rebalancing orders. Hereinafter, we call the preset threshold the "order quantity threshold."

We investigated the appropriate rebalancing trade method by changing the order quantity threshold. As a result, we observed that the number of rebalancing trades and the total amount of rebalancing orders can be reduced by increasing the order quantity threshold. Moreover, we can also confirm that the underlying market volatility increases as the amount of the managed underlying asset increases when the order quantity threshold is small.

## 2. Artificial Market Model

In view of the discussion of artificial market models in Introduction, for this study, we built a new artificial market model with an investor agent who manages leveraged assets on the basis of the artificial market models of Yagi et al. [14]. This choice was made because Yagi et al. [14] built a simple agent and pricing mechanism which could reproduce the statistical characteristics of the kinds of long-term price fluctuations observed in empirical analyses, to investigate the relationship among four market liquidity indicators by changing the parameters. Hereafter, we call the investor agent, who manages leveraged assets with the aim of achieving a rate of return several times greater than that of the underlying asset, a leveraged ETF agent. Therefore, there are two types of agents in our model, namely, a normal agent as a general investor and a leveraged ETF agent, as described in the following subsections. Specifically, there are $n$ normal agents, agents $j = 1, \ldots, n$, and one leveraged ETF in our market.

In the proposed model, only one risk asset (i.e., the underlying asset of a leveraged ETF) is available for trading. The mechanism for determining the price in this model is a continuous double auction (continuous trading session) [14]. It means that if there are some sell (buy) order prices in the order book that are lower (higher) than the agent's buy (sell) order price, then the agent's order is immediately matched to the lowest sell order (highest buy order) in the order book. Hereafter, we call this a "market order." If there are no such orders in the order book, then the order does not match any other order and remains in the order book. The remaining orders in the order book are cancelled time $t_c$ (order effective period) after the order was placed. The tick size, which is the minimum unit for the price, is $\delta P$. An agent can possess assets indefinitely because the quantity of cash of the agent is set indefinitely. Agents can also short sell.

*2.1. Order Process.* As described above, there are two types of investor agents: a normal agent [14] and a leveraged ETF agent.

A normal agent selected at random sends an order whose quantity is one. However, the same agent will not order more than once until all normal agents have ordered. A leveraged ETF agent places a rebalancing order as needed as each normal agent places an order. The time $t$ is incremented by 1 whenever a normal agent places an order except when a rebalancing trade is not successful. Thus, the process moves one step forward even when a trade of a normal agent does not occur and this new order is placed on the order book.

The order prices of normal agent $j$ by transaction are determined as shown below. The rate of change of the price expected by agent $j$ at time $t$ (the expected return) $r_{ej}^t$ is given by

$$r_{ej}^t = \frac{1}{w_{1j}^t + w_{2j}^t + u_j} \left( w_{1j}^t r_{e1,j}^t + w_{2j}^t r_{e2,j}^t + u_j \epsilon_j^t \right), \quad (1)$$

where $w_{ij}^t$ is the weight of the $i$-th term for agent $j$ and is set according to the uniform distribution between 0 and $w_{i,\max}$ at the start of the simulation and then varied by using the learning process described later. Furthermore, $u_j$ is the weight of the third term and is set according to the uniform distribution between 0 and $u_{\max}$ at the start of the simulation and kept constant thereafter. We assume that the degree of the three trading strategies, which are fundamental strategy, technical strategy, and noise trading, will be spread across agents, as we will explain in detail later. Thus, we model the trading weights as random variables independently chosen.

Equation (1) is composed of the three trading strategies: fundamental strategy, technical strategy, and noise trading. The reason why these strategies are implemented in our model is that many empirical studies found that the fundamental strategy, technical strategy, or both were used generally for any market and any time (e.g., Menkhoff and Taylor [15]). We also implement noise trading to model objectively investors' desire for a better strategy through trial and error. The meaning of the term fundamental strategy is as follows. The risk assets have their own fundamental prices measured based on related economic and financial factors. In theory, the market price is equal to the fundamental price. However, in the actual market, the market price is not always equal to the fundamental price. Fundamentalists who are fundamental strategy investors predict that the market price is going to converge to the fundamental price in the future and buy (sell) the stocks when the market price is lower (higher) than the fundamental price. The meaning of the term technical strategy is as follows. A technical strategy focuses on patterns of price movements, trading signals, and various other analytical charting tools to evaluate a stock's strength or weakness. There are many technical strategies in actual markets. We implement the trend following strategy, which buys the asset when its price trend goes up and sells when its trend goes down, expecting price movements to continue, because Iihara et al. [16] showed that many investors including institutional investors use the trend following strategy in their empirical analysis.

The initial term in equation (1) normalizes the impact of the three trading strategies. The first term in parentheses on the right-hand side of equation (1) represents the fundamental strategy, which indicates that an agent expects a positive (negative) return when the market price is lower (higher) than the fundamental price. The term $r_{e1,j}^t$ is the expected return of the fundamental strategy for agent $j$ at time $t$, given by $r_{e1,j}^t = \ln(P_f/P^{t-1})$, where $P_f$ is the fundamental price, which is constant over time, and $P^t$ is the market price at time $t$. The market price is set to the most recent price at the time if no trading is occurring. The initial market price is set to the fundamental price, i.e., $P^0 = P_f$.

The second term represents the technical strategy, which indicates that an agent expects a positive (negative) return when the historical return is positive (negative). Here, $r_{e2,j}^t$ is the expected return of the technical strategy for agent $j$ at time $t$, given by $r_{e2,j}^t = \ln(P^{t-1}/P^{t-1-\tau_j})$, where $\tau_j$ is set according to the uniform distribution between 1 and $\tau_{\max}$ at the start of the simulation.

The third term represents the noise strategy. Here, $\epsilon_j^t$ is a normally distributed random error with mean zero and standard deviation $\sigma_\epsilon$.

Based on the expected return $r_{ej}^t$, the expected price $P_{ej}^t$ is found using the following equation:

$$P_{ej}^t = P^{t-1} \exp(r_{ej}^t). \quad (2)$$

The order price $P_{oj}^t$ is set according to the uniform distribution between $P_{ej}^t - P_d$ and $P_{ej}^t + P_d$, where $P_d$ is constant.

The choice between buying and selling is determined by the relative sizes of the expected price $P_{ej}^t$ and the order price $P_{oj}^t$.

(i) An agent places a buy order for one share if $P_{ej}^t > P_{oj}^t$

(ii) An agent places a sell order for one share if $P_{ej}^t < P_{oj}^t$

On the other hand, the leveraged ETF agent places an order when the amount of the rebalancing trade required by the calculation method as described below exceeds the order quantity threshold $V_{\text{thr}}$.

Assume that $\text{NAV}^t$ and $L^t$ are the net asset value and the actual leverage ratio of the leveraged ETF agent at time $t$, respectively. Then $\text{NAV}^t$ and $L^t$ are defined as follows:

$$\begin{aligned} \text{NAV}^t &= P^{t-1} S^t + C^t, \\ L^t &= \frac{P^{t-1} S^t}{\text{NAV}^t}, \end{aligned} \quad (3)$$

where $S^t$ and $C^t$ are the number of the underlying assets and the amount of cash that the leveraged ETF agent possesses at time $t$, respectively. The leveraged ETF has to achieve a rate of return the target leverage ratio of the ETF $L$ times greater than that of the underlying asset. Therefore, the leveraged ETF agent's trading starts under conditions that satisfy $L$. In other words, the initial number of the underlying assets is set as $S^0 = LC^{\text{init}}/P^0$, where $C^{\text{init}}$ is the initial cash, and the initial cash after the leveraged ETF is set as $C^0 = C^{\text{init}} - P^0 S^0 = C^{\text{init}} - P^0 LC^{\text{init}}/P^0 = (1-L)C^{\text{init}}$. Note that the leveraged ETF agent borrows cash and buys the underlying asset when it cannot satisfy the target leveraged ratio of the ETF. The initial NAV of the leveraged ETF agent $\text{NAV}^0$ is defined as $\text{NAV}^0 = P^0 S^0 + C^0$.

The amount of the rebalancing trade $V^t$ is calculated as follows [11]:

$$V^t = \frac{\lfloor \{(L-1)P^{t-1}S^t + L^t C^t\} \rfloor}{P^{t-1}}. \quad (4)$$

The rebalancing trade of the leveraged ETF agent is performed when the absolute value of $V^t$ exceeds the order quantity threshold $V_{\text{thr}}$. When the sign of $V^t$ is positive, the leveraged ETF agent sends a sell order. When the sign of $V^t$ is negative, the leveraged ETF agent sends a buy order. These orders are market orders because the leveraged ETF agent has to trade so that $L^t$ becomes equal to $L$.

*2.2. Learning Process.* We modelled the learning process as follows based on Mizuta et al. [12]. The reason why the learning process is necessary and why the process should be modelled as described below is given in Mizuta et al. [12].

For $r_{ei,j}^t$, learning is performed by each agent immediately before the agent places an order. That is, when $r_{ei,j}^t$ and $r_l^t = \ln(P^{t-1}/P^{t-1-t_l})$, where $r_l^t$ is the return for the agents' learning process, are the same sign, the value of $w_{i,j}^t$ is updated to $w_{i,j}^t + k_l |r_l^t| q_j^t (w_{i,\max} - w_{i,j}^t)$, where $k_l$ is a constant and $q_j^t$ is set according to the uniform distribution between 0 and 1. When $r_{ei,j}^t$ and $r_l^t$ have opposite signs, the value of $w_{i,j}^t$ is updated to $w_{i,j}^t - k_l |r_l^t| q_j^t w_{i,j}^t$.

Separately from the process for learning based on past performance, $w_{i,j}^t$ is reset with a small probability $m$,

according to the uniform distribution between 0 and $w_{i,\max}$, to model objectively investors' desire for a better weight through trial and error. On the other hand, the learning process of the leveraged ETF agent is not modelled, because the strategy of the leveraged ETF agent depends on the size of the tracking error.

## 3. Simulation

*3.1. Overview.* In this paper, we consider the rebalancing trade method that has as little influence on the underlying asset market volatility as possible while maintaining small tracking errors. In this section, we conduct experiments to find a reasonable rebalancing trade method by changing the order quantity threshold. We also analyse the impact of the leveraged ETF agent's net asset value (NAV) on the underlying asset market by changing the initial cash of the leveraged ETF agent.

The initial cash of the leveraged ETF agent $C^{\text{init}}$ is defined as $C^{\text{init}} = 1{,}000{,}000 \times C_{\text{mag}}$. We express the NAV of the leveraged ETF agent by changing the initial cash coefficient $C_{\text{mag}}$. We set $C_{\text{mag}}$ at each of 10, 20, 30, 40, 50, 60, 70, 80, 90, and 100. The reason why we implemented $C_{\text{mag}}$ in our model is to control the relative amount of the initial cash $C^{\text{init}}$. By changing $C_{\text{mag}}$, we examine the effect of the rebalancing trades of the leveraged ETFs with different initial cash amounts on the underlying asset market. The initial cash of the leveraged ETFs in actual markets corresponds to $C^{\text{init}}$, not $C_{\text{mag}}$.

The order quantity threshold $V_{\text{thr}}$ is set according one of five patterns: 1, 2, 3, 4, and 5 when the initial cash coefficient $C_{\text{mag}}$ is 10 and proportionally larger for larger values of $C_{\text{mag}}$, as shown in Table 1.

Here, we introduce $V_{\text{nor}}$, which is calculated as the order quantity threshold divided by the initial cash coefficient (i.e., $V_{\text{thr}}/C_{\text{mag}}$), for our discussion of the patterns of the order quantity threshold. In the experimental environment, for a given value of $V_{\text{nor}}$, the results indicate that the order quantity threshold increases as the initial cash coefficient increases, so there is no need to consider the effect of the initial cash amount on the rebalancing trade.

The order quantity threshold $V_{\text{thr}}$ decides the minimum order quantity per rebalancing trade. The rebalancing trade is performed when the order quantity per rebalancing trade is larger than $V_{\text{thr}}$. By changing $V_{\text{thr}}$, we attempt to confirm the effect of the order quantity per rebalancing trade and the number of rebalancing trades on the underlying asset market. However, as $C^{\text{init}}$ increases, the order quantity per rebalancing trade also increases. If $V_{\text{thr}}$ is used as the order quantity threshold in the experiments, the number of rebalancing trades inevitably increases when $C^{\text{init}}$ is large. Therefore, we introduced $V_{\text{nor}}$ as $V_{\text{thr}}$ divided by $C_{\text{mag}}$, so that the number of rebalancing trades does not depend on the size of $C^{\text{init}}$. Finally, by changing $V_{\text{nor}}$, we examine the effect of the order quantity per rebalancing trade and the number of rebalancing trades on the underlying asset market. Note that $V_{\text{thr}}$ corresponds to the minimum order quantity in the market, if the leveraged ETF manager sets a minimum order quantity per rebalancing trade and performs the rebalancing trade.

TABLE 1: Order quantity threshold ($V_{\text{thr}}$) for each set of conditions.

| | | $V_{\text{nor}}$ | | | | |
|---|---|---|---|---|---|---|
| | | 0.1 | 0.2 | 0.3 | 0.4 | 0.5 |
| $C_{\text{mag}}$ | 10 | 1 | 2 | 3 | 4 | 5 |
| | 20 | 2 | 4 | 6 | 8 | 10 |
| | 30 | 3 | 6 | 9 | 12 | 15 |
| | 40 | 4 | 8 | 12 | 16 | 20 |
| | 50 | 5 | 10 | 15 | 20 | 25 |
| | 60 | 6 | 12 | 18 | 24 | 30 |
| | 70 | 7 | 14 | 21 | 28 | 35 |
| | 80 | 8 | 16 | 24 | 32 | 40 |
| | 90 | 9 | 18 | 27 | 36 | 45 |
| | 100 | 10 | 20 | 30 | 40 | 50 |

From the above, $V_{\text{nor}}$ is set to one of five values, 0.1, 0.2, 0.3, 0.4, or 0.5, and $C_{\text{mag}}$ is set to 10, 20, 30, 40, 50, 60, 70, 80, 90, or 100. Simulations were performed for all combinations, and simulation results each represent the averages of 100 simulation runs.

We set the initial values of the model parameters as follows. $n = 1{,}000$, $w_{1,\max} = 1$, $w_{2,\max} = 5$, $u_{\max} = 1$, $\tau_{\max} = 15{,}000$, $\sigma_\epsilon = 0.03$, $P_d = 1{,}000$, $t_c = 10{,}000$, $t_l = 10{,}000$, $k_l = 4$, $m = 0.01$, $\delta P = 1$, $P_f = 10{,}000$, $t_{\max} = 1{,}000{,}000$, and $L = 2.0$.

*3.2. Validation of Proposed Artificial Market.* As many empirical studies have mentioned [17, 18], a fat tail and volatility clustering appear in actual markets, which are two stylized facts of financial markets. Therefore, we set the artificial market parameters so as to replicate these features.

Table 2 shows the statistics for stylized facts in the case that the initial cash coefficient is 10 for which we calculated the price returns, i.e., $\ln(P^t/P^{t-1})$, at intervals of 100 time units. As shown, both kurtosis and autocorrelation coefficients for squared returns with several lags are positive, which means that the runs for all five patterns replicate a fat tail and volatility clustering. This indicates that the model replicates long-term statistical characteristics observed in real financial markets. Since similar results were obtained for the other values of the initial cash coefficient, those results are omitted here due to space limitations.

A previous empirical study showed that kurtosis of returns and autocorrelation coefficients for squared returns with several lags are positive when a fat tail and volatility clustering appear in actual markets [19]. For example, the kurtoses of monthly log returns of U.S. bonds, the S&P composite index of 500 stocks, and Microsoft are 4.86, 7.77, and 1.19, respectively. Note that kurtosis of returns depend on the kinds of financial assets. Likewise, autocorrelation coefficients for squared returns with several lags of the S&P composite index of 500 stocks are 0.0536, 0.0537, 0.0537, 0.0538, and 0.0538 when lags are from 1 to 5, respectively. It seems that the results of our model do not differ significantly from these results. Therefore, it can be confirmed that the proposed model is valid.

*3.3. Simulation Results.* We observed the number of rebalancing trades, the total rebalancing order quantity, and the rebalancing order quantity per trade, for which results are shown in Tables 3–5, respectively. For these tables and Table 6 discussed below, entries with dashes indicate that the market collapsed, so that the stylized facts of the market were not obtained and the corresponding value could not be measured.

Table 3 shows that the number of rebalancing trades decreases as $V_{nor}$ increases and $C_{mag}$ increases. Table 4 indicates that the total rebalancing order quantity decreases as $V_{nor}$ increases, but increases as $C_{mag}$ increases. The rebalancing order quantity per trade increases as $V_{nor}$ increases and $C_{mag}$ increases, as shown in Table 5.

Table 6 shows that the market volatility, specifically the standard deviation of market price volatility, decreases as $V_{nor}$ increases. On the other hand, the market volatility increases as $C_{mag}$ increases when $V_{nor}$ is small, whereas when $V_{nor}$ is large, it does not change significantly. Note that market collapses occur when $C_{mag}$ is large and $V_{nor}$ is small.

## 4. Discussion

In this section, first, we discuss what constitutes an appropriate rebalancing trading method for leveraged ETF managers, meaning a trading method that maintains the actual leverage ratio at the target leverage ratio of the ETF with a small rebalancing order quantity. Next, we discuss the impact of the leveraged ETF agent's NAV on the underlying market. Finally, we attempt to suggest what rebalancing trade method would be most appropriate based on those market impacts.

*4.1. Impact of the Order Quantity Threshold on the Market.* For the leveraged ETF managers, it is preferable to keep the actual leverage ratio at the target leverage ratio of the ETF with a small rebalancing order quantity. Therefore, we will discuss how changes in the order quantity threshold affect the rebalance trading. Here, we will fix $C_{mag}$ and discuss the impact of changes in $V_{nor}$.

The results in Table 4 indicate that the total rebalancing order quantity decreases as $V_{nor}$ increases. In other words, Table 4 shows that the rebalancing order quantity can be reduced by increasing the order quantity threshold. The reason for this can be intuitively understood because the results in Table 4 are a combination of the results in Tables 3 and 5, but are discussed in more detail below.

The number of rebalancing trades decreases as $V_{nor}$ increases is shown in Table 3. This phenomenon is thought to depend on the magnitude of the change in market price from the previous period when the rebalancing order quantity is equal to the order quantity threshold. The magnitudes of change in market price are shown in Table 7. For example, the $V_{nor}$ 0.1 column of Table 7 indicates that a rebalancing trade is performed when the market price changes ± 5 from the previous rebalancing trade price. As a result, when the market price increases from 10,000 to 10,005, a rebalancing trade is performed for $V_{\text{nor}}$ equal to 0.1, though it is not performed if $V_{\text{nor}}$ is equal to 0.2. If the price goes on to increase further to 10,010, then a rebalancing trade is performed both when $V_{\text{nor}}$ is 0.1 and when $V_{\text{nor}}$ is 0.2. As a result, the number of rebalancing trades is 2 when $V_{\text{nor}}$ is 0.1 and 1 when $V_{\text{nor}}$ is 0.2. Thus, the number of rebalancing trades decreases as $V_{\text{nor}}$ increases.

Table 2: Stylized facts. $C_{mag} = 10$.

| | | $V_{nor}$ | | | | |
|---|---|---|---|---|---|---|
| | | 0.1 | 0.2 | 0.3 | 0.4 | 0.5 |
| Kurtosis | | 8.93 | 7.26 | 4.18 | 2.54 | 2.13 |
| | Lag | | | | | |
| | 1 | 0.23 | 0.22 | 0.17 | 0.16 | 0.13 |
| Autocorrelation coefficients for squared returns | 2 | 0.19 | 0.17 | 0.12 | 0.11 | 0.10 |
| | 3 | 0.17 | 0.15 | 0.11 | 0.08 | 0.07 |
| | 4 | 0.15 | 0.11 | 0.09 | 0.07 | 0.05 |
| | 5 | 0.12 | 0.10 | 0.06 | 0.05 | 0.04 |

Table 3: Number of rebalancing trades.

| | | $V_{nor}$ | | | | |
|---|---|---|---|---|---|---|
| | | 0.1 | 0.2 | 0.3 | 0.4 | 0.5 |
| $C_{mag}$ | 10 | 6,785 | 2,152 | 705 | 248 | 100 |
| | 20 | 5,286 | 1,788 | 616 | 218 | 91 |
| | 30 | 4,855 | 1,531 | 564 | 201 | 86 |
| | 40 | — | 1,400 | 504 | 187 | 83 |
| | 50 | — | 1,290 | 463 | 179 | 81 |
| | 60 | — | — | 440 | 178 | 79 |
| | 70 | — | — | 424 | 173 | 79 |
| | 80 | — | — | 425 | 169 | 80 |
| | 90 | — | — | 421 | 176 | 81 |
| | 100 | — | — | 433 | 182 | 87 |

Table 4: Total rebalancing order quantity.

| | | $V_{nor}$ | | | | |
|---|---|---|---|---|---|---|
| | | 0.1 | 0.2 | 0.3 | 0.4 | 0.5 |
| $C_{mag}$ | 10 | 12,851 | 5,595 | 2,337 | 1,028 | 508 |
| | 20 | 26,515 | 10,727 | 4,366 | 1,849 | 926 |
| | 30 | 41,780 | 14,557 | 6,176 | 2,569 | 1,318 |
| | 40 | — | 18,395 | 7,400 | 3,187 | 1,692 |
| | 50 | — | 21,401 | 8,517 | 3,806 | 2,058 |
| | 60 | — | — | 9,717 | 4,583 | 2,420 |
| | 70 | — | — | 10,859 | 5,151 | 2,844 |
| | 80 | — | — | 12,430 | 5,744 | 3,292 |
| | 90 | — | — | 13,782 | 6,721 | 3,731 |
| | 100 | — | — | 15,524 | 7,705 | 4,471 |

Table 5: Rebalancing order quantity per trade.

| | | $V_{nor}$ | | | | |
|---|---|---|---|---|---|---|
| | | 0.1 | 0.2 | 0.3 | 0.4 | 0.5 |
| $C_{mag}$ | 10 | 1 | 2 | 3 | 4 | 5 |
| | 20 | 4 | 5 | 6 | 8 | 10 |
| | 30 | 8 | 8 | 10 | 12 | 15 |
| | 40 | — | 12 | 14 | 16 | 20 |
| | 50 | — | 16 | 17 | 20 | 25 |
| | 60 | — | — | 21 | 25 | 30 |
| | 70 | — | — | 25 | 29 | 35 |
| | 80 | — | — | 28 | 33 | 40 |
| | 90 | — | — | 32 | 37 | 45 |
| | 100 | — | — | 35 | 41 | 50 |

Table 6: Underlying market volatility ($\times 10^{-3}$).

|  |  | $V_{\text{nor}}$ | | | | |
|---|---|---|---|---|---|---|
|  |  | 0.1 | 0.2 | 0.3 | 0.4 | 0.5 |
| $C_{\text{mag}}$ | 10 | 1.35 | 0.95 | 0.81 | 0.79 | 0.78 |
|  | 20 | 1.78 | 1.03 | 0.82 | 0.78 | 0.78 |
|  | 30 | 2.27 | 1.07 | 0.81 | 0.79 | 0.78 |
|  | 40 | — | 1.12 | 0.81 | 0.77 | 0.77 |
|  | 50 | — | 1.16 | 0.80 | 0.76 | 0.77 |
|  | 60 | — | — | 0.79 | 0.76 | 0.76 |
|  | 70 | — | — | 0.78 | 0.74 | 0.75 |
|  | 80 | — | — | 0.78 | 0.73 | 0.74 |
|  | 90 | — | — | 0.78 | 0.73 | 0.74 |
|  | 100 | — | — | 0.78 | 0.72 | 0.73 |

Table 7: Magnitude of change from previous market price when rebalancing order quantity is equal to order quantity threshold.

|  | $V_{\text{nor}}$ | | | | |
|---|---|---|---|---|---|
|  | 0.1 | 0.2 | 0.3 | 0.4 | 0.5 |
| Price change | 5 | 10 | 15 | 20 | 25 |

The results in Table 5 show that the rebalancing order quantity per trade increases as $V_{\text{nor}}$ increases. However, the number of rebalancing trades decreases rapidly as $V_{\text{nor}}$ increases, as shown in Table 3. Thus, the total rebalancing order quantity decreases when $V_{\text{nor}}$ increases.

### 4.2. Impact of the Leveraged ETF Agent's NAV on the Market.

The rebalancing order quantity is also affected by the leveraged ETF agent's NAV. In this section, we will discuss the impact on the underlying market by changing the leveraged ETF agent's NAV. Here, we discuss the impact of changes in $C_{\text{mag}}$ with $V_{\text{nor}}$ fixed.

Table 6 confirms that market volatility increases as $C_{\text{mag}}$ increases when $V_{\text{nor}}$ is small. However, if $C_{\text{mag}}$ increases too much, the market will collapse. Therefore, the following findings were obtained: rebalancing trades of leveraged ETFs having large NAVs have a significant impact on the market in the sense of increased market volatility. In particular, the increase in market volatility becomes significant when rebalancing transactions with a small order volume are frequently performed.

The reason for the above can be explained by the following two points. One is that the rebalancing order quantity per trade greatly exceeds the order quantity threshold for large $C_{\text{mag}}$, especially when $V_{\text{nor}}$ is small, as can be seen by comparing Tables 1 and 5. The other is that the market volatility increases as the rebalancing order quantity per trade increases.

The former reason can be explained using the following example. When $V_{\text{nor}}$ is 0.1, regardless of the size of $C_{\text{mag}}$, if the magnitude of change in market price exceeds 5, then a rebalancing trade occurs. The order quantities at this time calculated based on equation (3) are 1 and 3 when $C_{\text{mag}}$ is 10 and 30, respectively. These values are consistent with the order quantity threshold for both of these $C_{\text{mag}}$ values. However, if the change in market price is ±9, the quantities of orders for rebalancing trade are 1 and 5 from equation (6) when $C_{\text{mag}}$ is 10 and 30, respectively. In the case of $C_{\text{mag}} = 30$, the order quantity is larger than the order quantity threshold. Thus, as $C_{\text{mag}}$ increases, the actual order quantity per rebalancing trade eventually becomes greater than the order quantity threshold. This is more likely to occur as $V_{\text{nor}}$ is smaller. This is because as $V_{\text{nor}}$ becomes small, the magnitude of change in market price from the previous period when the rebalancing order quantity equal to the order quantity threshold also becomes small, so the actual order quantity more easily exceeds the order quantity threshold. On the other hand, as $V_{\text{nor}}$ becomes large, the magnitude of change in market price from the previous period when the rebalancing order quantity equal to the order quantity threshold also becomes large, so the actual order quantity is less likely to exceed the order quantity threshold.

Next, the reason why market volatility increases as the rebalancing order quantity per trade greatly exceeds the order number threshold is described. As the leveraged ETF agent's orders are market orders, once the rebalancing trade orders are placed, the trade always takes place, and the best quote in the order book for the market is always removed. As $C_{\text{mag}}$ increases, the quantity of each order becomes larger, and the market price after the rebalancing trade is higher from the previous price. As a result, market volatility can be considered to have increased.

### 4.3. Suggestion for the Appropriate Rebalancing Trade Method.

As can be seen from the above discussion, increasing $V_{\text{nor}}$ reduces the total rebalancing order quantity and also reduces market volatility. This finding also holds when $C_{\text{mag}}$ is large, as long as the market does not collapses. In other words, when the mechanism for determining the price is the continuous double auction, it is possible to reduce the total rebalancing order quantity and the impact on the price formation by making several trades with a large rebalancing order quantity rather than making many more trades with small rebalancing order quantities. This result suggests that increasing the minimum number of orders in a rebalance trade can reduce the impact of such a rebalancing trade on the market and leads to keeping the actual leverage ratio at the target leverage ratio of the ETF.

## 5. Conclusions and Future Work

In this study, we investigated what rebalancing trading method has the least influence on the underlying market volatility while maintaining small tracking errors by using an artificial market. The following findings were obtained. First, increasing the order quantity threshold reduces the total rebalancing order quantity and decreases market volatility. Next, increasing the NAV of the leveraged ETF agent causes increased market volatility, and in particular, the market volatility significantly increases as the order quantity threshold becomes smaller, although it does not change significantly when the order quantity threshold is large. From these findings, we conclude that increasing the minimum number of orders in a rebalancing trade defines an appropriate trading method from the perspective of the leveraged ETF manager, as well as that of the underlying market.

Our future work can be summarized as follows. In this study, by investigating the trading method focusing on the effect of rebalancing trade on the underlying market, it was found that it is effective to increase the order quantity threshold for the rebalancing trades. However, a larger-order quantity threshold results in a longer period during which a tracking error occurs. Therefore, we should discuss how to avoid such a situation. Furthermore, when the order quantity threshold is large, the market volatility does not change significantly even if the NAV of the leveraged ETF increases. Therefore, we intend to investigate the reason for this.

## Data Availability

All data used to support the findings of this study are included within the article.

## Disclosure

## Conflicts of Interest

The authors declare that they have no conflicts of interest.